# The Effect of Declustering on the Size Distribution of Mainshocks

Leila Mizrahi[1], Shyam Nandan[1], Stefan Wiemer[1]

[1]Swiss Seismological Service, ETH Zurich, Switzerland

Corresponding author: Leila Mizrahi, Sonneggstrasse 5, 8092 Zürich, Switzerland

leila.mizrahi@sed.ethz.ch

December 16, 2020

## KEY POINTS:

- We find that the choice of declustering method substantially influences the earthquake-size distribution of mainshocks.

- We show using simulation-based tests that declustering can introduce a systematic bias to the size distribution of earthquakes, potentially biasing hazard assessment.

ABSTRACT

Declustering aims to divide earthquake catalogs into independent events (mainshocks), and dependent (clustered) events, and is an integral component of many seismicity studies, including seismic hazard assessment. We assess the effect of declustering on the frequency-magnitude distribution of mainshocks. In particular, we examine the dependence of the $b$-value of declustered catalogs on the choice of declustering approach and algorithm-specific parameters. Using the catalog of earthquakes in California since 1980, we show that the $b$-value decreases by up to 30% due to declustering with respect to the undeclustered catalog. The extent of the reduction is highly dependent on the declustering method and parameters applied. We then reproduce a similar effect by declustering synthetic earthquake catalogs with known $b$-value, which have been generated using an Epidemic-Type Aftershock Sequence (ETAS) model. Our analysis suggests that the observed decrease in $b$-value must, at least partially, arise from the application of the declustering algorithm on the catalog, rather than from differences in the nature of mainshocks versus fore- or aftershocks. We conclude that declustering should be considered as a potential source of bias in seismicity and hazard studies.

INTRODUCTION

Models for probabilistic seismic hazard analysis (PSHA), e.g. (Petersen et al., 2018; Pace et al., 2006; Wiemer et al., 2009a; Gerstenberger et al., 2020), are commonly based on the approach described by Cornell (1968), which assumes earthquake occurrence times to be representable by a stationary Poisson process. The long-term seismicity rate in a region is considered to be constant in time, reflecting a constant deformation rate and hence constant energy input at any given location, driven by plate tectonics. In reality, earthquakes trigger aftershocks, which in turn trigger their aftershocks, and so on, leading to intense clustering of earthquakes in space and time (Ogata, 1998; Jackson and Kagan, 1999; Helmstetter and Sornette, 2003). Earthquakes can also occur in swarms (Hainzl and Fischer, 2002; Hainzl, 2004), lasting days to months, sometimes comprising thousands of earthquakes in one location, which are followed by long periods of quiescence. Consequently, the recorded earthquake catalogs, especially modern instrumental ones that are complete down to small magnitudes, always show conspicuous deviations from Poissonianity. Average seismicity rates in regions with recent large sequences are therefore not representative of the long-term seismic hazard, indicating a potentially substantial location-dependent bias of seismicity rates.

## Aims and challenges of declustering

So-called *declustering algorithms* aim to divide earthquake catalogs into clusters of dependent events and retain only the independent event of each such cluster. Although Luen and Stark (2012) find that Poissonianity depends on *"the declustering method, the catalog, the magnitude range, and the statistical test"*, it is generally assumed that a properly declustered earthquake catalog satisfies the condition of being Poissonian (Gardner and Knopoff, 1974, van Stiphout et al., 2012). Because of the requirement of Poissonianity for the current approach to PSHA, rate estimation for hazard assessment is often done on the basis of declustered catalogs (Field et al., 2014; Petersen et al., 2018; Pace et al., 2006; Wiemer et al., 2009a, Drouet et al., 2020; Meletti et al., 2017; Akinci et al., 2018; Sesetyan et al., 2018; Beauval et al., 2013; Waseem et al., 2019; Woessner et al., 2015). In this sense, PSHA approaches estimate mainshock rates rather than total seismicity rates.

While Poissonianity of the declustered catalog is necessary for a declustering method to serve its purpose, this condition does not ensure a unique solution to the declustering problem. To avoid inadvertently rewarding the excessive removal of events from the catalog, an additional criterion is required. However, as the actual triggering processes are not currently known and nature does not provide us with labels such as *'mainshock', 'aftershock', 'foreshock',* or *'swarm member',* we lack an objective criterion for the performance evaluation of declustering methods. Several algorithms have been proposed and used in the past (Gardner and Knopoff, 1974; Gruehthal, 1985; Uhrhammer, 1986; Reasenberg, 1985; Zaliapin et al., 2008; Zhuang et al., 2002); Marsan and Lengline, 2008; see van Stiphout et al. (2012) for an overview.

## Effects of declustering on PSHA

In a study on the effect of declustering on hazard results for the city of Istanbul, Azak et al. (2018) found that peak ground acceleration values vary by up to 20% depending on the declustering method. Marzocchi and Taroni (2014) discuss the need for declustering for PSHA, concluding that it is only necessary to avoid a bias in the spatial distribution of earthquake occurrences. Furthermore, considering that aftershocks can also cause considerable damage, they find that the neglecting of aftershock effects due to declustering may lead to significant underestimation of seismicity rates and hence of seismic hazard. In this regard, Iervolino et al. (2018) and Iervolino (2019) have proposed a generalization of the hazard integral to re-introduce aftershock hazard in PSHA. Moreover, van Stiphout et al. (2011) found that the choice of declustering method has a major effect on seismicity rate-change estimations. On the other hand, sensitivity studies to different declustering approaches in Switzerland have shown that the impact of declustering on the hazard is often negligible (Wiemer et al., 2009b). The need for, potential biases introduced by, and alternatives to declustering have also been discussed in the context of seismicity forecasting (see Nandan et al., 2019a; Schorlemmer and Gerstenberger, 2007). In particular, the issue is raised that a mainshock forecast can only be tested against a mainshock 'truth' which is inherently dependent on the somewhat arbitrary choice of declustering method, yielding full seismicity forecasts the only objectively testable type of forecast.

## Effects of declustering on the $b$-value

A major role in the calculation of seismicity rates is played by the $b$-value of the empirical Gutenberg-Richter (GR) law (Gutenberg and Richter, 1944), which describes the frequency distribution of earthquake magnitudes. Typically, $b$-values of earthquake catalogs lie close to $1$ (Kagan, 1999; Kamer and Hiemer, 2015), but have been found to vary with time, region, depth, and stress regime. Several studies have also reported higher $b$-values during swarms or in volcanic areas (Main et. al., 1992; Henderson et. al., 1992; Frohlich and Davis, 1993; Wiemer and Wyss, 1997; Schorlemmer et. al., 2005; Petruccelli et. al., 2019; Wyss et al., 1997). Kagan (1999), Kamer and Hiemer (2015), and Marzocchi et al. (2020) discussed a variety of potential technical causes of $b$-value variations, such as magnitude binning, network coverage, catalog incompleteness, or the finiteness of data. Moreover, imposing a GR law on declustered catalogs, as is commonly done in seismic hazard analysis, often results in a significantly lower $b$-value compared to full catalogs (Kagan, 2010; Christophersen et. al., 2011; Field et. al., 2014; Petersen et. al., 2018). Some argue that this behavior is a property naturally inherent to mainshocks (Knopoff, 2000). On a similar note, Gulia et al. (2018) suggested that the $b$-value of typical aftershock sequences is on average 20% higher than the mainshock $b$-value, and that this increase in $b$-value is a long-lasting effect for several years.

However, it is debatable whether the $b$-value of a declustered catalog is at all meaningful. Most declustering methods define a mainshock as the largest event of an independent cluster. If one assumes a GR law-type Pareto distribution of magnitudes on the full catalog, one should not at the same time assume a GR law-type Pareto distribution of mainshock magnitudes. The distribution of the maximum of a set of independent and identically distributed random variables, i.e., the distribution of mainshock magnitudes, can be derived from fundamental principles of probability theory (Kolmogoroff, 1934). Lombardi (2003) gave a mathematical description of how mainshock magnitude distribution follows from the full-catalog GR law. She showed that the difference in $b$-value between mainshocks and all events becomes minimal when a corrected log-likelihood function is used in the maximum likelihood estimation of the mainshock $b$-value. The mainshock magnitude distribution she proposed depends on the empirical distribution of cluster sizes emerging from the declustering process. Given this result and assuming that different declustering algorithms will lead to different cluster size distributions, it is expected that different declustering methods will also lead to different mainshock magnitude distributions. Hence, $b$-values of mainshocks, when estimated in the usual way, are expected to be biased as an artifact of declustering.

Similarly, Zhuang and Ogata (2006) found that the magnitude distribution of mainshocks defined via the Epidemic-Type Aftershock Sequence (ETAS) model (see their article for the definition or Text S4 in the electronic supplement to this article for details on the ETAS model) departs from the GR law and that the full catalog $b$-value is valid for mainshocks in the asymptotic case where $m \rightarrow \infty$. For lower magnitude mainshocks, one could argue that a GR law with lower $a$- and $b$-values than those of the full catalog presents an acceptable approximation of the true, non-Pareto distribution of mainshock magnitudes. However, when the logarithms of the numbers $N(m)$ and $N_{\text{main}}(m)$ of earthquakes and mainshocks of magnitude $M > m$ are both described by linear terms of the form

$$log_{10} N(m) = a - b \cdot m, \tag{1}$$

$$log_{10} N_{\text{main}}(m) = a_{\text{main}} - b_{\text{main}} \cdot m, \tag{2}$$

where $b_{main} \neq b$, the two lines intersect at a point

$$m_x = \frac{a - a_{main}}{b - b_{main}}. \tag{3}$$

If $b_{main} < b$, this means that the expected number of mainshocks of magnitude $M > m_x$ is larger than the expected number of total earthquakes of magnitude $M > m_x$, even though the observed number of mainshocks can never be larger than the observed number of earthquakes.

## Paper outline

Considering the importance of the $b$-value for seismicity studies and seismic hazard estimates, we here systematically assess the influence of declustering on mainshock size distribution. To do so, we first verify that imposing a GR law on mainshocks yields a $b$-value that does indeed depend on the choice of the declustering method applied. Then, we show that a similar effect is observed for synthetic catalogs with known $b$-value, whose magnitude distribution by design does not distinguish mainshocks and other events. Furthermore, we illustrate the consequences of approximating mainshock magnitude distribution with a Pareto distribution and calculate the tipping point magnitude $m_x$, above which the bias introduced by declustering cannot be interpreted as mainshock-specific behavior.

The rest of the paper is organized as follows. In DATA, we describe the earthquake catalog used for this study and discuss the coupled estimation of completeness magnitude and $b$-value. In METHOD, we describe the declustering methods and corresponding parameter choices. There, we also describe the ETAS model, which is used for the simulation of synthetic catalogs and furthermore serves as the basis for two of the declustering methods. We then present and discuss our main results in RESULTS AND DISCUSSION and state our CONCLUSION. The supplemental material to this article contains a more detailed description of all methods and algorithms used.

Moreover, it contains analyses of the sensitivity of full catalog $b$-value and mainshock $b$-value on the completeness magnitude $m_c$.

DATA

In this study, we use the ANSS Comprehensive Earthquake Catalog (ComCat) provided by the U.S. Geological Survey (see DATA AND RESOURCES) with 'preferred' magnitudes as defined in ComCat, in the collection area around the state of California as in the RELM testing center (Schorlemmer and Gerstenberger, 2007). The choice of the study region is motivated mainly by the high seismicity in the area and by completeness at low magnitudes of the catalog for several decades (see Hutton et al., 2006), both ensuring that a large and representative amount of data can be used in our study. We consider events of magnitude $M \geq 0.0$, with magnitudes rounded into bins of size $\Delta M = 0.2$. Figure S1 in the electronic supplement to this article shows that the $b$-value is insensitive to bin size for reasonable choices of $m_c$. It also shows that $b(m_c)$ is more stable for $\Delta M = 0.2$ compared to $\Delta M = 0.1$. The time frame used is January 1, 1970 until September 30, 2019, of which only the events on or after January 1, 1980 are used for the estimation of $b$-values. We subsequently call this set of events the *incomplete primary catalog*. The earlier events make up the *incomplete auxiliary catalog*. As earthquake clusters may occur close to the start of the primary time period, ignoring auxiliary events could lead to unwanted deficiencies in cluster detection and mainshock identification (Wang et al., 2010; Schoenberg et al., 2010; Nandan et al., 2019b). Our choice of time periods aims to achieve balance between long enough primary and auxiliary periods, and low completeness magnitude in the primary catalog thanks to seismic network configuration (see e.g. Hutton et al., 2006).

### $b$-Value Estimation and Completeness Magnitude

Estimating the $b$-value of a catalog requires knowledge of its completeness magnitude $m_c$, the magnitude threshold above which all events are assumed to be detected. Assuming too low values for $m_c$ can cause severe underestimation of the $b$-value (see Figure S1 in the electronic supplement to this article). On the other hand, assuming overly conservative values for $m_c$ leads one to discard a large portion of the data, making $b$-value estimates imprecise. In reality, $m_c$ is not known and has to be estimated itself. Several methods to do so have been proposed; see Mignan and Woessner (2012) for an overview. Commonly, $m_c$ is estimated by defining it as the magnitude threshold above which earthquakes follow the GR law. In this sense, the estimation of $b$-value and $m_c$ becomes a coupled problem; one cannot be estimated without knowledge of the other. In Texts S1 and S2 and Figure S2 in the electronic supplement to this article, we adapt the method proposed by Clauset et al. (2009) to jointly estimate $m_c$ and $b$-value, ultimately arriving at a value of 3.6 for $m_c$. A sensitivity analysis (see Figure S3 in the electronic supplement to this article) shows that the results presented in the following sections are insensitive to reasonable choices of $m_c$.

Setting the value of $m_c$ to 3.6 implies that we subsequently use the subset of events with magnitude $M \geq 3.6$ of the previously described (binned) catalog. This filter is applied to both the incomplete primary and the incomplete auxiliary catalog, yielding the *(complete) primary* and *(complete) auxiliary catalog*. Figure 1(a) and (c) show the spatial and temporal distribution of events in the catalog with magnitude $M \geq 3.6$, where auxiliary events are highlighted in yellow. For our primary catalog we obtain a $b$-value of 1.01, as illustrated in Figure 1(b).

## METHOD

To better understand the influence of declustering on the $b$-value, we first apply five often used declustering techniques with different parameter and window choices to the same real catalog. We then apply the same declustering methods, with standard parameters, to a set of 2000 synthetic catalogs. The synthetic catalogs are generated using a basic ETAS model (see Ogata, 1998; Veen et and Schoenberg, 2008; Nandan et al., 2017; Nandan et al., 2019a), which is described in Text S4 in the electronic supplement to this article. Table 1 shows the parameters used in the simulation of synthetic catalogs. They were obtained by applying expectation maximization (Veen et and Schoenberg, 2008; Nandan et al., 2017) to the primary and auxiliary California catalog, to support the comparability of real and synthetic catalogs. For a detailed description of the ETAS model, as well as the algorithms used for inversion and simulation, see Text S4 in the electronic supplement to this article. We use the synthetic catalogs to test whether declustering introduces any systematic bias to the mainshock size distribution. As in the case of synthetic catalogs, the distribution from which magnitudes are drawn is known and is assumed to be the same for mainshocks and aftershocks, any changes in $b$-value observed after declustering must have their origin in the application of declustering algorithms.

To further understand the consequences of approximating mainshock magnitude distribution with a lower-$b$-value GR law, we compare the ratio

$$r(m) = \frac{N_{\mathrm{main}}(m)}{N(m)} \qquad (4)$$

of mainshocks among earthquakes of magnitude $M > m$ between observation and approximation, for the different declustering methods with standard parameter settings applied to the Californian catalog. We calculate $m_x$ (see Equation 3), above which $r(m) > 1$ implies that the introduced bias can impossibly be supported by observations.

We examine the declustering methods proposed by Reasenberg (1985), Zaliapin et al. (2008), and window methods as proposed by Gardner and Knopoff (1974), Gruehthal (1985), and Uhrhammer (1986). We also consider two versions of declustering based on the ETAS model (Zhuang et al., 2002). For the detailed description of all declustering algorithms and parameter ranges applied, see Texts S3 and S4 and Tables S1 to S3 in the electronic supplement to this article; we give a short description of each method below. Note that the non-parametric stochastic declustering algorithm proposed by Marsan and Lengline (2008) is not used here. This is because of its similarity to the already considered parametric stochastic declustering alternative provided by the ETAS model. The main difference to ETAS declustering is that the triggering rate, described as $g(t, x, y, m)$ in Text S4 in the electronic supplement to this article, is there obtained empirically, without presuming the laws (S12)-(S14). In their analysis of southern California seismicity, they observe that their empirically derived triggering rate follows laws similar to those described in (S12)-(S14), which, likewise, were originally discovered empirically.

### Short descriptions of the declustering methods applied in this article

1. *Reasenberg* (1985) introduced an algorithm that has been used in numerous studies and recent PSHA, e.g. in Ecuador (Beauval et al., 2013) or Afghanistan (Waseem et al., 2019). It defines earthquake interaction zones in space and time. Here, we apply the spatial interaction relationships proposed by Reasenberg (1985), and Wells and Coppersmith (1994), and the parameter ranges for temporal interaction zones recommended by Schorlemmer and Gerstenberger (2007).

2.  *Window methods*, as first described by Gardner and Knopoff (1974) use space-time windows around large events to identify their fore- and aftershocks. Different formulations of such window boundaries have been suggested and are applied in this study (Gardner and Knopoff, 1974; Gruenthal, 1985; Uhrhammer, 1986; see van Stiphout et al., 2012). We use the original formulation by Gardner and Knopoff (1974) as the standard window, which is used in the Uniform California Earthquake Rupture Forecast (UCERF3, Field et al., 2014). Generally, window methods are widely used in modern regional and national seismic hazard models, see Drouet et al. (2020) for France, Meletti et al. (2017) for Italy, Sesetyan et al. (2018) for Turkey, Woessner et al. (2015) for Europe (ESHM13).

3.  *Zaliapin*'s (2008) alternative approach applies a Gaussian mixture model on space-time nearest-neighbor distances between events to distinguish independent from dependent events.

4.  The *ETAS* model is used here in two ways. Firstly, it is used to simulate synthetic earthquake catalogs upon which declustering methods are applied to study their effects. Secondly, the ETAS model induces an alternative, parametric approach to declustering, which was introduced by Zhuang et al. (2002). We consider two versions of declustering based on the ETAS model, which differ in their definition of mainshocks and are described in detail in Text S4 in the electronic supplement to this article. *'ETAS-Main'* defines the largest event of a cluster to be the mainshock, while *'ETAS-Background'* defines events to be mainshocks if they are not triggered. The definition used in ETAS-Background is in the spirit of the ETAS model, where background earthquakes of any size can trigger cascades of aftershocks. ETAS-Main, on the other hand, imposes the mainshock definition used in the other methods, in the interest of comparability. We subsequently call those methods which define mainshock as the largest events *'mainshock methods'*. Note that because of its different definition of mainshocks, ETAS-Background is unsuited to be applied in the standard PSHA approach, which is designed to work with mainshock methods.

## RESULTS AND DISCUSSION

### The Disparity Between Declustering Methods

The cumulative number of mainshocks for different declustering methods with standard parameter and window choices, compared to the full California catalog, is shown in Figure 2(a). The diversity among the resulting declustered catalogs is remarkable. Mainshock rates vary by a factor of 6.1 between the most and least 'aggressive' algorithm. Moreover, while the removal of temporal clusters is the primary goal of the declustering process, some are still clearly visible after declustering with Reasenberg's method, and still recognizable, though less pronounced, after applying Zaliapin's method. Gardner-Knopoff and ETAS (Main and Background) appear to be more successful at achieving temporal Poissonianity.

The observed and fitted complementary cumulative frequency functions (CCFFs) of mainshock magnitudes are shown in Figure 2(b) and (c). Observed absolute frequencies of large events ($M \geq 6.4$) are somewhat similar for all declustering methods, with the exception of ETAS-Background. Relative frequencies of large events versus small events vary strongly between methods, which manifests itself in slope differences between the CCFFs. Note that for the mainshock methods, the aggressiveness of the method coincides with the extent of slope decrease. This effect can be explained as a consequence of the methods' mainshock definition. Since small events are less likely to be identified as mainshocks, they are more likely to be removed from the catalog, increasing relative frequencies of large events. ETAS-Background, in contrast, does not seem to preferentially remove events from specific magnitude ranges.

Figure 2(d) illustrates the consequences of estimating seismic hazard based on a mainshock GR law with lower $b$-value. For the California catalog, observed and approximated evolutions of $r(m) = \frac{N_{\text{main}}(m)}{N(m)}$ are shown for mainshocks obtained by Gardner-Knopoff declustering. The magnitude $m_x$ (see Equation 3) above which rate overestimation cannot be denied, is given also for Reasenberg, Zaliapin and ETAS-Main declustered catalogs (see Figure S4 in the electronic supplement to this article for the corresponding plots). $m_x$ varies between 6.9 and 8.8, where higher values of $m_x$ are predominantly observed for declustering methods which do not succeed at achieving Poissonianity in time. Furthermore, most declustering methods show considerable deviations of observed $r(m)$ from its approximation already at lower magnitudes. For instance, the approximation of $r_{6.6} = r(6.6)$ lies between 0.72 and 0.86, depending on mainshock definition, even though all definitions except ETAS-Main classified *all* $M > 6.6$ events to be mainshocks. Note that ETAS-Background is excluded from this part of the analysis due to its inapplicability in the standard PSHA approach.

## $b$-Value of Declustered Catalogs

### *Observations on real data*

The relative frequency increase of large events translates into a lower $b$-value when a GR law is imposed on the frequency-magnitude distribution of mainshocks. In Figure 3(a), $b$-value is plotted against $a$-value of the declustered California catalog, comparing the effects of varying declustering methods and parameters. We find that $b$-values of declustered catalogs vary strongly with declustering algorithms. Values between 0.73 and 1.00 are attained without any significant gap. A general trend is recognizable among the mainshock methods: removal of more events correlates with lower $b$-values, indicating a penchant of these methods to relatively remove more smaller events than larger ones. The $b$-value obtained with ETAS-Background does not significantly differ from the full catalog $b$-value. These observations are in line with the explanation given above, which describes the $b$-value decrease as a consequence of the mainshock definition, and are expected knowing the results by Lombardi (2003), Zhuang and Ogata (2006), Kagan (2010) and van Stiphout et al. (2011). A sensitivity analysis of the $b$-value to the completeness magnitude $m_c$ (see Figure S3 in the electronic supplement to this article) shows that the $b$-value decrease after declustering is an effect that is observed regardless of the reasonable choice of $m_c$, with the extent of the decrease being characteristic of each method.

*Observations on simulated data*

Synthetic catalogs, where all magnitudes are drawn from one single distribution, show lower $b$-values after declustering. In Figure 3(b), the distribution of mainshock $b$-values of 2000 ETAS-simulated catalogs is shown for different declustering methods with standard parameter settings, aligned according to the median observed $a$-value of the respective method. The mainshock $b$-values of the same methods applied to the California catalog are indicated with stars; error bars mark the estimated standard error. If no declustering, or ETAS-Background declustering, is applied, the estimated $b$-value of synthetic catalogs is consistent with the $b$-value used in their simulation. At the same time, $b$-values of synthetic catalogs declustered with mainshock methods are always lower than the $b$-value used in their simulation. Comparing the extent of the effect across different declustering methods, synthetic and real data have the same qualitative behavior. Similarly, the $a$-value decrease is observed to be method-characteristic.

The effect of declustering on the $b$-value is more pronounced in synthetic data, for all methods. A possible explanation for this is that all declustering methods assume isotropic spatial distribution of aftershocks, which is known to be wrong in reality, but valid for synthetic catalogs. Hence, cluster detection is facilitated in synthetic catalogs, resulting in more effective removal of small events compared to the real catalog.

## Remarks Regarding ETAS Declustering

1. Despite ETAS being the generative process of synthetic catalogs, a large difference in $b$-value is observed after ETAS-Main declustering. This is not a flaw in the generative process or the parameter inversion. On the contrary, this behavior is expected. The low $b$-value of ETAS-Main-declustered catalogs is due to the imposed definition of 'mainshock' as the largest event of a cluster, not to be confused with ETAS' notion of background events. This concept of mainshocks is not relevant in the generation of catalogs. Imposing such a definition leads to selective removal of small events rather than to the removal of aftershocks in the true ETAS sense. With ETAS, aftershocks are temporally restricted to occur after their triggering events but may have larger magnitudes.

2. Declustering with ETAS-Background allows a comparison between the $b$-value of background events according to the ETAS definition and the full catalog. No significant difference is observed, both in the case of synthetic and real catalogs. The difference only arises when the rule of maximum magnitude is applied.

3. For synthetic data, the underlying branching structure is known by design of the experiment. In contrast, cluster detection for real data requires the lengthy process of inversion and probabilistic cluster assignment. Thus, compared to synthetic data, cluster detection is intrinsically less correct for real data, and declustering is inclined to be less effective. It is reasonable to assume that this circumstance explains the particularly pronounced difference in $b$-value in the case of ETAS-Main declustering.

ETAS simulations do not distinguish the magnitude distribution of mainshocks versus aftershocks. A difference in pre- and post-declustering $b$-value of a declustered catalog that was generated using ETAS can, therefore, only have its origin in the systematic selection of large events as mainshocks. The purely declustering-induced and strongly method-dependent decrease in $b$-value suggests that other potential causes, such as a different nature of mainshocks compared to fore- or aftershocks, have negligible effects on the mainshock $b$-value. While the possibility cannot be precluded that a part of the effect is due to the change in stress state before and after major events (e.g. Gulia et al., 2018), the notably arbitrary effect of declustering should not be ignored. The mere observation of a lower $b$-value of mainshocks is no proof of its meaningfulness; the observation of artifactual effects of declustering on the mainshock $b$-value, however, is a reason to doubt its meaningfulness.

## CONCLUSION

We demonstrate that a decrease in overall $b$-value of the California catalog after declustering is observed for a variety of declustering methods and parameter settings. Furthermore, the extent of the decrease is highly dependent on the algorithm applied. A general trend is observed, suggesting that more 'aggressive' algorithms tend to be accompanied by a more pronounced $b$-value decrease, ETAS-Background being the only exception to this rule. With a medial resulting $a$-value among the methods considered, it leaves the $b$-value unchanged. Finally, we find that all the above-described effects can be reproduced in synthetic data, which was generated using a constant $b$-value for all events.

Our results indicate that declustering substantially affects the earthquake size distribution. Imposing a GR law on declustered catalogs, therefore, leads to $b$-values which are biased to a somewhat arbitrary and not immediately apparent extent. This bias leads to an overestimation of seismic hazard above a certain magnitude $m_x$. Thus, we can conclude that the current state of practice of equating seismic hazard with mainshock rates which follow a GR law can be accused of three deficiencies. One is the non-verifiability of any mainshock definition. Secondly, fore- or aftershocks can be large and devastating. Neglecting aftershock effects may give rise to a substantial underestimation of seismic hazard (Marzocchi and Taroni, 2014). And lastly, the earthquake size distribution resulting from the procedure causes hazard overestimation for events above a certain size.

One may argue that increasing the relative frequency of large events and decreasing the absolute frequency of all events have antagonistic effects on the absolute frequency of large events, justifying any choice of declustering method. Indeed, most hazard studies ignore previous findings and continue to calculate hazard in the usual way. However, we believe that two wrongs do not make a right. To be precise, two wrongs make a right only for one particular magnitude, $m_x$. Our analysis suggests that above $m_x$, classical hazard studies overestimate the seismic hazard, whereas below $m_x$, they underestimate it. While the accusation of underestimation could partially be rejected by insisting that declustering reveals the true mainshocks and that aftershock effects are deliberately excluded from the scope of hazard assessment, we have shown that the overestimation cannot be similarly attributed to a true $b$-value that is revealed by declustering, but that this resulting $b$-value is biased to a non-negligible extent.

It is crucial to be aware of this issue when estimating seismic hazard. While analysis of earthquake dependency is inevitable to eliminate spatial bias for the calculation of seismicity rates (Marzocchi and Taroni, 2014), basing calculations solely on declustered catalogs is not an appropriate approach. One alternative possibility is to use ETAS models to assess seismicity rates (see e.g. Field et al., 2015). In a pseudo-prospective forecasting experiment on the Californian catalog conducted by Nandan et al. (2019a), ETAS models generally outperform all competing smoothed seismicity models and models based on strain rates. Using hundreds of thousands of simulations of possible scenarios as the basis for a forecast, they intrinsically account for the spatiotemporal clustering of earthquakes. This approach incorporates the non-Poissonian nature of reality while reducing the spatial bias encountered in undeclustered catalogs. At the same time, ETAS relies only on the GR law of the full catalog and therefore avoids making assumptions on the frequency-magnitude distribution of somewhat arbitrarily selected large events.

Other ways to address this matter may exist. What is essential is to recognize the problematic aspects of doing hazard assessment based on declustered catalogs and to find a way to address the issues presented here.

## DATA AND RESOURCES

The ANSS Comprehensive Earthquake Catalog (ComCat) provided by the U.S. Geological Survey was searched using https://earthquake.usgs.gov/data/comcat/ (last accessed on November 30, 2019). The catalog used can be found online: link to the catalog.

The electronic supplement to this article contains a more detailed description of all methods and algorithms used. Moreover, it contains analyses of the sensitivity of full catalog $b$-value and mainshock $b$-value on the completeness magnitude $m_c$. Finally, observed and approximated ratio of mainshocks among earthquakes of magnitude $M > m$ is shown using different declustering methods for mainshock definition.

## ACKNOWLEDGMENTS

The authors wish to thank Celso Reyes for providing python implementations of the declustering algorithms for Reasenberg, Zaliapin, and window method declustering, as well as Laurentiu Danciu, Arnaud Mignan, as well as two anonymous reviewers for their helpful feedback on an earlier version of this manuscript. This work has received funding from the ETH research grant for "Enabling dynamic earthquake risk assessment (DynaRisk)" and from the European Union's Horizon 2020 research and innovation programme under grant agreement No.821115, Real-Time Earthquake Risk Reduction for a Resilient Europe (RISE).

FULL MAILING LIST FOR EACH AUTHOR

Leila Mizrahi, Sonneggstrasse 5, 8092 Zürich, Switzerland, leila.mizrahi@sed.ethz.ch

Shyam Nandan, Sonneggstrasse 5, 8092 Zürich, Switzerland, snandan@ethz.ch

Stefan Wiemer, Sonneggstrasse 5, 8092 Zürich, Switzerland, stefan.wiemer@sed.ethz.ch

TABLES

**Table 1:** ETAS parameters used for catalog simulation, obtained by expectation maximization.

| Parameter | Value |
|---|---|
| $\log(k_0)$ | $-2.49$ |
| $a$ | $1.69$ |
| $\log(c)$ | $-2.95$ |
| $\omega$ | $-0.03$ |
| $\log(\tau)$ | $3.99$ |
| $\log(d)$ | $-0.35$ |
| $\gamma$ | $1.22$ |
| $\rho$ | $0.51$ |
| $\log(\mu)$ | $-7.17$ |

LIST OF FIGURE CAPTIONS

**Figure 1:** Earthquake catalog used in this analysis. (a) Seismicity map. Dots represent earthquakes in the catalog with $M \geq 3.6$, where dot size indicates magnitude. Events between 1970 and 1980, which serve as auxiliary data, are marked in yellow. Solid black line marks the California state boundary, dotted line marks the boundary of the considered region. (b) Absolute frequency distribution of magnitudes above and below $m_c$ (black versus grey diamonds). Solid black line shows the GR law fitted to the catalog of events with $M \geq 3.6$. (c) Temporal distribution of the events shown in (a), with identical size and color coding.

**Figure 2:** Properties of the California catalog of mainshocks larger than or equal to M3.6, depending on declustering method. Standard parameter settings (and standard window) of each method are used for declustering. (a) Cumulative number of mainshocks. Dotted black line represents the full (non-declustered) catalog. The rapid increase in seismicity highlighted in circles corresponds to the 2010 ($M_w$ 7.2) El Mayor-Cucapah earthquake in Baja California, Mexico. (b) Empirical Complementary Cumulative Frequency Function (CCFF, diamonds) of mainshock magnitudes. Empty black diamonds represent the full California catalog. The two lines are the fitted CCFF for Gardner-Knopoff declustered and full catalog. (c) Fitted CCFF for declustered catalogs compared to full catalog GR law fit. Fitted b-values are given. (d) Observed (diamonds) and approximated (line) evolution of $r(m) = \frac{N_{main}(m)}{N(m)}$ for the Gardner-Knopoff declustered catalog. Black dotted line marks $r(m) \equiv 1$. $m_x$ and $r_{6.6}$ are given, also for Reasenberg, Zaliapin and ETAS-Main declustered catalogs. Note the different x-axis for (d) compared to (b) and (c).

**Figure 3:** (a) $a$-value versus $b$-value of the declustered California catalog depending on declustering method and parameters. Each dot represents one variation of parameter settings, stars with error bars represent standard parameter settings. Marked with (W) are window methods. The dotted grey line and grey area indicate the $b$-value of the non-declustered catalog and its uncertainty. (b) Distribution of mainshock $b$-values of 2000 simulated catalogs, depending on declustering method (with standard parameter settings and standard (Gardner-Knopoff) window), plotted against median $a$-value per method. Stars with error bars represent the $a$- and $b$-value of the regional earthquake catalog from (a) for the respective methods. White dot, black box and black line represent median, interquartile range and adjacent values. The dotted line displays the $b$-value used for catalog generation, which corresponds to the full-catalog b-value observed in the Californian primary catalog.

FIGURES

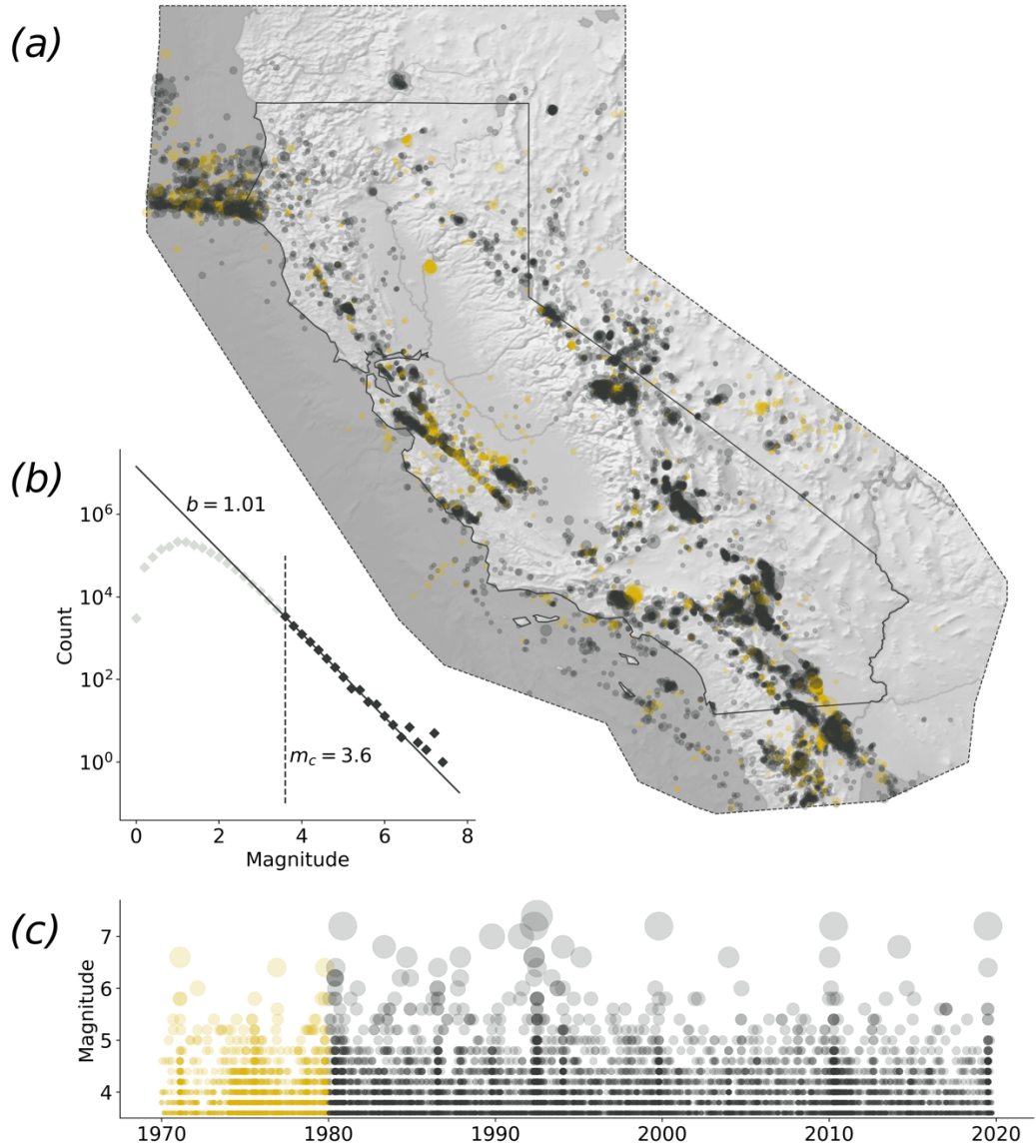

*(a)*

*(b)*

*(c)*

Figure 1: Earthquake catalog used in this analysis. (a) Seismicity map. Dots represent earthquakes in the catalog with $M \geq 3.6$, where dot size indicates magnitude. Events between 1970 and 1980, which serve as auxiliary data, are marked in yellow. Solid black line marks the California state boundary, dotted line marks the boundary of the considered region. (b) Absolute frequency distribution of magnitudes above and below $m_c$ (black versus grey diamonds). Solid black line shows the GR law fitted to the catalog of events with $M \geq 3.6$. (c) Temporal distribution of the events shown in (a), with identical size and color coding.

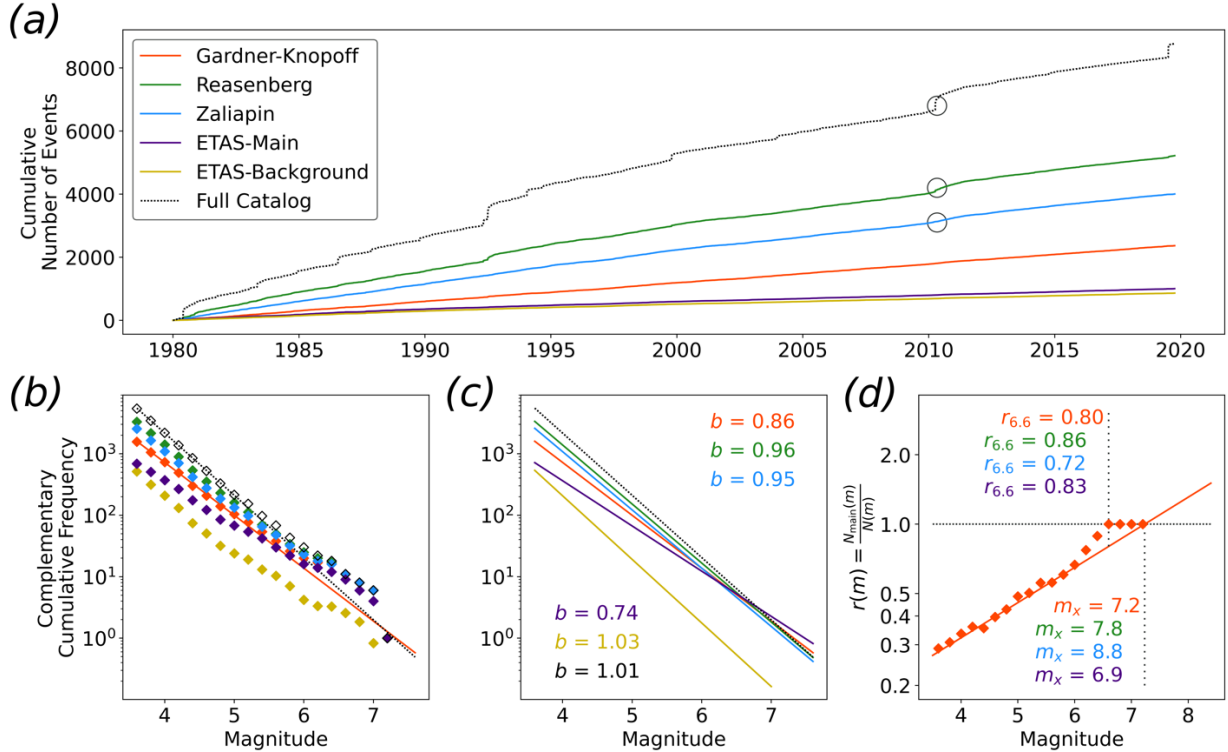

*Figure 2: Properties of the California catalog of mainshocks larger than or equal to M3.6, depending on declustering method. Standard parameter settings (and standard window) of each method are used for declustering. (a) Cumulative number of mainshocks. Dotted black line represents the full (non-declustered) catalog. The rapid increase in seismicity highlighted in circles corresponds to the 2010 ($M_w$ 7.2) El Mayor-Cucapah earthquake in Baja California, Mexico. (b) Empirical Complementary Cumulative Frequency Function (CCFF, diamonds) of mainshock magnitudes. Empty black diamonds represent the full California catalog. The two lines are the fitted CCFF for Gardner-Knopoff declustered and full catalog. (c) Fitted CCFF for declustered catalogs compared to full catalog GR law fit. Fitted b-values are given. (d) Observed (diamonds) and approximated (line) evolution of $r(m) = \frac{N_{main}(m)}{N(m)}$ for the Gardner-Knopoff declustered catalog. Black dotted line marks $r(m) \equiv 1$. $m_x$ and $r_{6.6}$ are given, also for Reasenberg, Zaliapin and ETAS-Main declustered catalogs. Note the different x-axis for (d) compared to (b) and (c).*

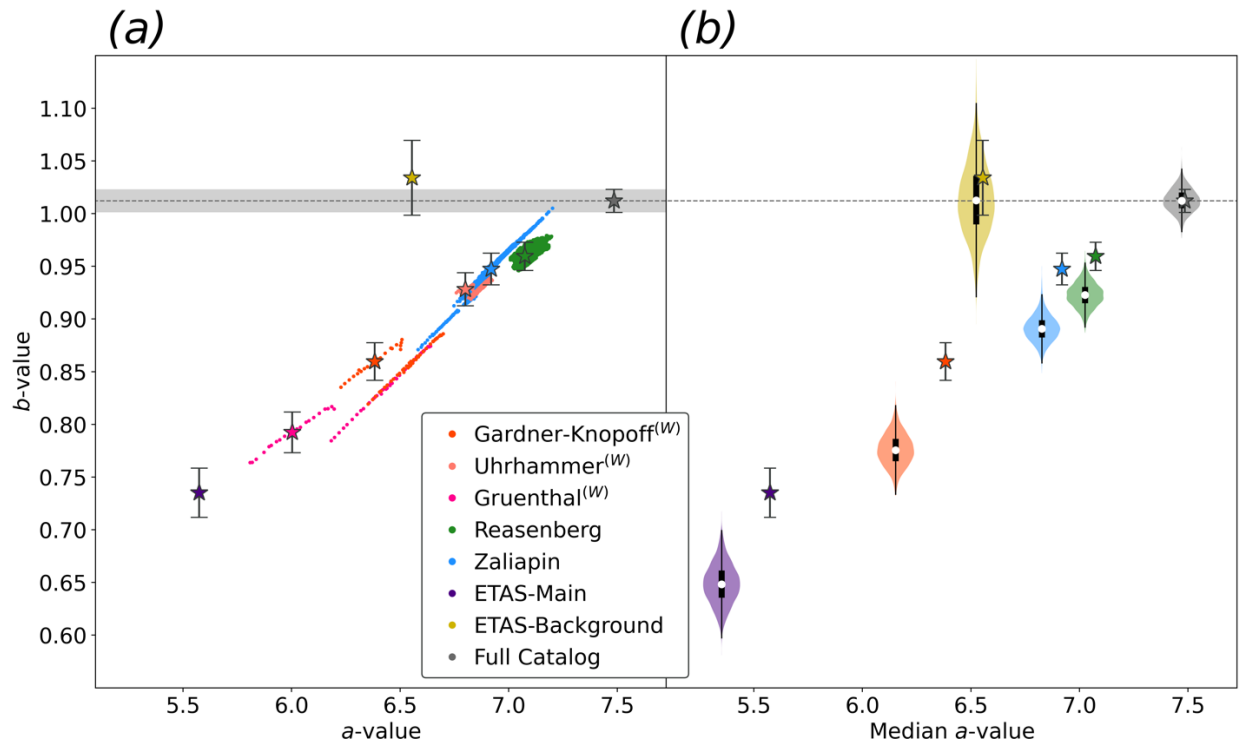

*Figure 3: (a) a-value versus b-value of the declustered California catalog depending on declustering method and parameters. Each dot represents one variation of parameter settings, stars with error bars represent standard parameter settings. Marked with (W) are window methods. The dotted grey line and grey area indicate the b-value of the non-declustered catalog and its uncertainty. (b) Distribution of mainshock b-values of 2000 simulated catalogs, depending on declustering method (with standard parameter settings and standard (Gardner-Knopoff) window), plotted against median a-value per method. White dot, black box and black line represent median, interquartile range and adjacent values. Stars with error bars represent the a- and b-value of the regional earthquake catalog from (a) for the respective methods. The dotted line displays the b-value used for catalog generation, which corresponds to the full-catalog b-value observed in the Californian primary catalog.*

Supplemental Material for

## The Effect of Declustering on the Size Distribution of Mainshocks

Leila Mizrahi, Shyam Nandan, and Stefan Wiemer

## Description of the Supplemental Material

This document contains a detailed description of the methodology, including the formula used for $b$-value estimation and the algorithm for joint estimation of $b$-value and completeness magnitude, the description of all declustering methods and parameter ranges applied, as well as a description of the ETAS model, parameter inversion, and catalog simulation algorithm applied. We analyze the sensitivity on the completeness magnitude $m_c$ of the full catalog and mainshock $b$-value, and sensitivity on magnitude binning of the full catalog $b$-value. Finally, observed and approximated ratio of mainshocks among earthquakes of magnitude $M > m$ is shown using different declustering methods for mainshock definition.

## List of Supplemental Table Captions

**Table S1.** Parameter ranges applied in the Reasenberg declustering algorithm. Standard parameter settings are given in bold characters.

**Table S2.** Parameter ranges and window types applied in the window declustering methods. Standard parameter settings and window are given in bold characters.

**Table S3.** Parameter ranges and window types applied in the window declustering methods. Standard window and parameter settings are given in bold characters.

## List of Supplemental Figure Captions

**Figure S1.** Sensitivity of $b$-value on completeness magnitude and magnitude bin size $\Delta M$.

**Figure S2.** Testing $m_c^i$ between 2.0 and 6.0. (a): For integer values of $m_c^i$, theoretical and observed CDFs are represented as dotted black and red lines, respectively. Light grey and dark grey areas mark 5th to 95th, and 25th to 75th percentile of 10,000 randomly sampled CDFs. Grey areas are barely visible for small $m_c^i$, because they are in agreement with the theoretical CDF. (b): Evolution of $p$-value with $m_c^i$. (c): Derivation of $p$-value for integer values of $m_c^i$. Histogram shows the frequency distribution of KS distances of 10,000 randomly sampled CDFs to theoretical CDF. Red vertical line indicates KS distance of observed to theoretical CDF. (d) Evolution of $b$-value with $m_c^i$, with uncertainties.

**Figure S3.** Sensitivity of mainshock $b$-value on completeness magnitude for different declustering methods.

**Figure S4.** Observed and approximated (diamonds and lines) evolution of $r(m) = \frac{N_{main}(m)}{N(m)}$ for the California catalog. Declustering method varies with panel, marked with (W) are window methods. Grey lines show realizations of $r(m)$ for different declustering parameters, colored line shows the realization for standard parameters, black line marks $r(m) \equiv 1$. Colored diamonds mark observed values for standard declustering parameters. $m_x$ and $r_{6.6}$ are given for standard declustering parameters.

## Text S1.

### *b*-Value Estimation

According to the Gutenberg-Richter law (Gutenberg and Richter, 1944), frequency-magnitude distribution can be described as

$$N(m) = N_0 \cdot e^{-\beta \cdot m} = 10^a \cdot 10^{-b \cdot m},$$

where $N(m)$ is the number of earthquakes of magnitude $M > m$, $N_0$ is the total number of earthquakes above magnitude $M = 0$. $a$ and $b$ will subsequently be called '$a$-value' and '$b$-value', respectively. Note that $\beta = ln(10) \cdot b$.

$\beta$ is estimated using the formula proposed in (Tinti and Mulargia, 1987) for binned magnitude values (in contrast to continuous values). It reads

$$\hat{\beta} = \frac{1}{\Delta M} \ln(p), \tag{S1}$$

where

$$p = 1 + \frac{\Delta M}{\hat{\mu} - m_c}, \tag{S2}$$

$m_c$ is the completeness magnitude, $\hat{\mu}$ is the mean observed magnitude and $\Delta M$ is the bin size in which earthquake magnitudes are given, in our case $0.2$.

## Text S2.

### Completeness Magnitude

Estimating the $b$-value of a catalog requires knowledge of its completeness magnitude $m_c$, the magnitude threshold above which all events are assumed to be detected. Assuming too low values for $m_c$ can cause severe underestimation of the $b$-value (see Figure S2). On the other hand, assuming overly conservative values for $m_c$ leads one to discard a large portion of the data, making $b$-value estimates imprecise. In reality, $m_c$ is not known and has to be estimated itself. Commonly, $m_c$ is estimated by defining it as the magnitude threshold above which earthquakes follow the GR law (Mignan and Woessner, 2012). In this sense, the estimation of $b$-value and $m_c$ becomes a coupled problem; one cannot be estimated without knowledge of the other. In the following section, we adapt the method proposed by Clauset et al. (2009) to jointly estimate $m_c$ and $b$-value.

For the estimation of completeness magnitude, we use all events above magnitude 0.0 in the primary time period, in the same collection area and using the same binning as described in Section 2 of the paper. For simplicity, $m_c$ is chosen to be constant in time and space, with the knowledge that this assumption does not generally hold. Short-term aftershock incompleteness after large events and modifications to the seismic network or analysis procedures will increase (and decrease) the completeness magnitude (Woessner and Wiemer, 2005; Schorlemmer et al., 2010). We test a range of possible completeness magnitudes $m_c^i$ between $2.0$ and $6.0$. For each choice of $m_c^i$, the $b$-value $b^i$ is calculated for the events above $m_c^i$, using the formula proposed by Tinti and Mulargia (1987), (see Text S1). This yields a discretized GR law and its corresponding cumulative distribution function (CDF). Simultaneously, the observed cumulative distribution function of events above $m_c^i$ is computed. Observed and theoretical CDF are then compared to each other using the Kolmogorov-Smirnov (KS) distance. We then estimate $p$-values $p^i$: the probability of observing a KS distance of $D_{m_c^i}$ or higher, under the assumption that the observed magnitudes were drawn from a discretized GR law with a $b$-value of $b^i$. This is done by generating 10,000 random samples of the size of the original sample. For each such sample, we compute the KS distance of observed CDF to the theoretical CDF they were drawn from. The $p$-value $p^i$ is then calculated as the fraction of KS distances that are larger than $D_{m_c^i}$. Finally, we accept the completeness magnitude $m_c^i$ if $p^i \geq 0.05$, and set the according $b$-value estimator to $b^i$. Note that this selection of $m_c^i$ based on $p^i$ is unlike Clauset's method, where $m_c^i$ is chosen such that the KS distance $D_{m_c^i}$ is minimized.

Since the KS distance highly depends on sample size and can hence fluctuates strongly with $m_c^i$, this approach is more robust than simply minimizing the KS distance. After applying the described algorithm to our incomplete primary catalog, we choose $m_c = 3.6$ with $b \approx 1.01$ and $p \approx 0.09$. Figure S1 illustrates the test for five cases of $m_c^i$. In Figure S1a, theoretical and observed CDFs are depicted as dotted black and red lines, respectively. Light grey and dark grey areas (barely visible for small $m_c^i$) mark the 5th to 95th, and 25th to 75th percentile of 10,000 randomly sampled CDFs. In Figure S1c, the histograms of KS distances of 10,000 randomly sampled CDFs are shown for the same five $m_c^i$. Red vertical lines mark $D_{m_c^i}$ values. In Figure S1b, the evolution of the $p$-value is shown for all $m_c^i$ participating in the test, while Figure S1d shows the corresponding $b$-values $b^i$. Note that $m_c^i = 3.0$ is the lowest value to pass the test. However, because completeness is strictly required for $b$-value calculation, we choose the more conservative value of 3.6, as it is the smallest $m_c^i$ whose $p$-value is not deceeded by $p$-values of later $m_c^i$.

**Text S3.**

### Detailed Description of Declustering Methods

For Reasenberg and Zaliapin's declustering methods as well as window methods, we use the python codes developed for the 2020 update of the European Seismic Hazard Model (ESHM20, Danciu et al., 2019) available as an open-source contribution to the seismic hazard modeler's toolkit of OpenQuake (Weatherill et al., 2014; Pagani et al., 2015). The implementation of the ETAS declustering algorithm is described separately.

Cluster detection is applied to the full time period including 10 years of auxiliary data, but only those mainshocks that fall into the primary time period are considered for further analysis such as $b$-value estimation. This ensures that aftershocks of events happening before the start of the primary time period are not mistakenly identified as mainshocks. With the exception of ETAS-Background, all methods define a mainshock to be the largest event of a cluster detected by the method. The set of all mainshocks, including events that form a cluster of size 1, make up the declustered catalog.

### Reasenberg

Reasenberg (1985) introduced an algorithm that has been used in numerous studies, e.g. in Ecuador (Beauval et al., 2013) or Afghanistan (Waseem et al., 2019). It defines earthquake interaction zones in space and time. Clusters are built by linking dependent events as follows. If an event lies in the interaction zone of another, the two events are linked. Linked events build a cluster. If an event is linked to an event belonging to a cluster, it is added to the cluster. If two events belonging to different clusters are linked, the two clusters are merged to form one cluster.

Spatial interaction zones are defined around the current as well as the largest event of an ongoing sequence. Their boundaries depend on magnitude, and on the interaction formula that is being used. The two interaction formulae applied here are

$$f_{R_{1985}}(m) = 0.011 \cdot 10^{0.4m}, \tag{S3}$$
$$f_{W\&C_{1994}}(m) = 0.01 \cdot 10^{0.5m}, \tag{S4}$$

where $f_{R_{1985}}$ and $f_{W\&C_{1994}}$ are relationships proposed by Reasenberg (1985), and an updated version using the scaling relationships by Wells and Coppersmith (1994), respectively.

Temporal interaction zones are defined based on a probabilistic Omori law approach. The look-ahead time, within which events are being linked, is calculated such that the probability of detecting the next event in the cluster is equal to the parameter $p$, assuming a lower cut-off magnitude, $xm_{eff}$, an increase of the lower cut-off magnitude during clusters to $xm_{eff} + x_k \cdot M$, where $M$ is the magnitude of the largest event of the current sequence. $r_{fact}$ represents the number of crack radii around each earthquake within which new events of the cluster are considered (van Stiphout et al., 2012). $\tau_{min}$ and $\tau_{max}$ are the minimum and maximum allowed look-ahead time.

Table S1 shows the parameter ranges applied in this study, with standard parameters in bold characters. The ranges and standard settings are adopted from (Schorlemmer and Gerstenberger, 2007). $xm_{eff}$ is set to be the completeness magnitude $m_c$.

## Window Methods

Window methods, as first described by Gardner and Knopoff (1974), are widely used in regional and national seismic hazard models, see Drouet et al., (2019) for France, Akinci et al., (2018) for Italy, Sesetyan et al., (2018) for Turkey, Field et. al., (2014) for California (UCERF3), Woessner et al., (2015) for Europe (ESHM13). They define space time windows depending on mainshock magnitude and denote events within the window of a large event as fore- or aftershocks of said mainshock. In contrast to Reasenberg's method, higher level aftershocks are not considered. Furthermore, large events become mainshocks by definition.

Spatial windows are circular around epicenters and vary with magnitude. Temporal windows for foreshocks are per default identical to their corresponding aftershock time window but can be proportionally shortened or extended by multiplication with a factor $p_{\tau_{foreshock}}$. Furthermore, there is the possibility to cap time window length at a maximum number of days $\tau_{max}$. A variety of formulae for the definition of space time window boundaries has been proposed. In our analysis, we use the following three variants.

1) Garnder-Knopoff Window (Gardner and Knopoff, 1974)

$$\Delta_s(m) = 10^{0.1238m + 0.983}[km] \tag{S5}$$

$$\Delta_t(m) = \begin{cases} 10^{0.5409m-0.547}, & m < 6.5 \\ 10^{0.032m+0.983}, & m \geq 6.5 \end{cases}[days] \tag{S6}$$

2) Gruenthal Window (Gruenthal, 1985; see van Stiphout et al., 2012)

$$\Delta_s(m) = e^{1.77 + \sqrt{0.037 + 1.02m}}[km] \tag{S7}$$

$$\Delta_t(m) = \begin{cases} |e^{-3.95 + \sqrt{0.62 + 17.32\,m}}|, & m < 6.5 \\ 10^{2.8 + 0.024m}, & m \geq 6.5 \end{cases}[days] \tag{S8}$$

3) Uhrhammer Window (Uhrhammer, 1986)

$$\Delta_s(m) = e^{-1.024 + 0.804m}[km] \tag{S9}$$

$$\Delta_t(m) = e^{-2.87 + 1.235m}[days] \tag{S10}$$

Table S2 shows the parameter ranges applied in this study, with standard window and parameters in bold characters.

## Zaliapin

An alternative approach to declustering was proposed in (Zaliapin et al., 2008). Space time distances between pairs of events are calculated as

$$\eta_{ij} = \begin{cases} \left(t_{ij}\right)^\theta \cdot \left(r_{ij}\right)^d \cdot 10^{-bm_i}, & t_{ij} > 0 \\ \infty, & t_{ij} \leq 0 \end{cases}, \tag{S11}$$

where $t_{ij} = t_i - t_j$ is the time difference between event $i$ and event $j$, $r_{ij}$ is the spatial distance between their epicenters, and $m_i$ is the magnitude of the earlier event? $d$ is the fractal dimension of epicenters, $\theta$ is an exponent used to weight temporal distance relative to spatial distance, and $b$ is the $b$-value of the Gutenberg Richter law.

For each event, its nearest neighbor with respect to this distance measure can be identified. Using a Gaussian mixture model, nearest-neighbor space time distances between events are then classified into two categories. Smaller distances are interpreted as distances between dependent events, and larger distances represent distances between independent events. The Gaussian mixture model yields two pairs of mean and standard deviation of distances; one for dependent events, and one for independent events. Nearest neighbor distances are classified as dependent event distances, if their likelihood under the dependent normal distribution is larger than under the independent normal distribution. Earthquake clusters are then defined as tree-like structures of dependent nearest-neighbors.

Table S3 shows the parameter ranges applied in this study, with standard parameters in bold characters. These ranges were chosen based on recommendations given in the paper introducing the algorithm.

The other two declustering methods are quasi a side product of the ETAS inversion and is therefore described separately.

## Text S4.

### ETAS Model

A basic epidemic-type aftershock sequence (ETAS, (Ogata, 1998)) model is used here in two ways. Firstly, it is used to simulate synthetic earthquake catalogs upon which declustering methods are applied to study their effects. Secondly, ETAS provides an alternative, parametric approach to declustering, which was introduced by Zhuang et al. (2002).

Remark: Note here the important distinction between independent events in the ETAS sense, and mainshocks in the declustering sense.

In ETAS, one distinguishes triggered events and events that are not triggered, so-called independent events. Strictly speaking, all earthquakes are triggered by some underlying physical processes. What we mean by "not triggered" in this context is that the event is unlikely to be triggered by another earthquake that was observed in the catalog. This can be because the triggering earthquake was too weak to be detected, or that other physical processes that are not captured in the model were its cause.

Independent events and their aftershocks are modelled as a marked self-exciting point-process as follows. While independent events are assumed to be uniformly distributed in time and space, the aftershock triggering process is modelled to follow three fundamental principles derived from empirical laws.

1) The Utsu aftershock productivity law (Utsu, 1970) describes the number of aftershocks of an event given its magnitude. It describes the aftershock productivity $p_{AS}$ of an event to be exponentially increasing with the magnitude $m$ of the triggering event.

$$p_{AS}(m) = K \cdot e^{a(m-m_c)}. \tag{S12}$$

2) Aftershock occurrence time is modelled by an exponentially tapered Omori kernel (Omori, 1894), i.e. aftershock occurrence rate after waiting time $t$ is proportional to

$$q_{AS}(t) = \frac{e^{-t/\tau}}{(t+c)^{1+\omega}}.$$  (S13)

The non-tapered Omori law only allows for exponents larger than 1 in the above denominator, which seems to be an unreasonably strict mathematical condition. This can be bypassed by using the tapered Omori law.

3) Aftershock location is assumed to be distributed isotropically around the epicenter of the triggering event and aftershock occurrence rate $r_{AS}$ decreases with distance as follows.

$$r_{AS}(x, y) = \frac{1}{\left((x^2 + y^2) + d \cdot e^{\gamma(m-m_c)}\right)^{1+\rho}},$$  (S14)

where $x$ and $y$ are spatial distances in $x$- and $y$-direction between two events. The assumption of isotropy is a simplification which is known to be wrong in reality, as aftershocks tend to occur along fault systems rather than in circles around their triggering events.

In summary, the aftershock triggering rate $g(m)$ is defined as

$$g(t, x, y, m) = \frac{k_0 \cdot e^{a(m-m_c)}}{\frac{(t+c)^{1+\omega}}{e^{-t/\tau}} \cdot \left((x^2 + y^2) + d \cdot e^{\gamma(m-m_c)}\right)^{1+\rho}},$$  (S15)

where $m$ is the magnitude of the triggering event, and $\sqrt{x^2 + y^2}$ and $t$ are the spatial and temporal distance between triggering and triggered event, respectively. $k_0, a, c, \omega, \tau, d, \gamma$ and $\rho$ are constants. Nandan et al. (2019a) show that models allowing these parameters to vary in space, as proposed in (Nandan et al., 2017), significantly outperform the forecasting ability of models where they are spatially homogeneous. However, they also show that spatially homogeneous ETAS outperforms three declustering based smoothed seismicity models (SSMs), a simple SSM based on undeclustered data and a model based on strain rate data. For our purpose, spatially homogeneous ETAS is an adequate choice, as our main focus lies on the overall distribution of magnitudes, and not on spatial variations in seismicity. For analogous reasons, possible temporal variability of ETAS parameters is neglected in this analysis.

Furthermore, the constant $\mu$ represents the (spatially and temporally) uniform background intensity. The occurrence rate of earthquakes at a given time $t$ and place $(x, y)$ is then described as

$$\lambda(x, y, t) = \mu + \sum_{i: t_i < t} g(t - t_i, x - x_i, y - y_i, m_i)$$  (S16)

where $t_i, m_i, (x_i, y_i)$ are time, magnitude, and location of the $i^{th}$ event. Essentially, the occurrence rate $\lambda$ at a location $(x, y)$ at time $t$ is the sum of the independent earthquake rate and aftershock rates of all events preceding time $t$.

### ETAS Inversion

In order to obtain reasonable estimates for the above-mentioned ETAS parameters, one needs to solve an inversion problem: which set of parameters best describes the observed catalog? In this analysis, we use a so-called Expectation Maximization (EM) algorithm, as proposed by (Veen and Schoenberg, 2008). They find that, compared to commonly used maximum likelihood estimation, using EM for the inversion of ETAS parameters has substantial advantages in terms of convergence, bias, and robustness to the choice of starting values.

Starting with a set of randomly chosen initial parameters, we repeatedly undergo the expectation step E and the maximization step M, until a convergence criterion is met.

In the expectation step, probabilities $p_{ij}$ of each event $e_j$ to be triggered by each other event $e_i$ are calculated, given the current parameter estimates. Earthquakes in the auxiliary catalog can only serve as triggering events, not as triggered events, while events of the primary catalog can assume both roles. For any given target event, triggering probabilities are proportional to aftershock occurrence rates $g_{ij}$. The probability of each event $e_j$ to be an independent event (independence probability, $p_j^{ind}$) is proportional to the background rate. The normalization factor is chosen such that all triggering probabilities and the independence probability sum up to 1.

$$p_{ij} = \frac{g_{ij}}{\mu + \sum_{k \text{ s.t. } t_k < t_j} g_{kj}},$$  (S17)

$$p_j^{ind} = \frac{\mu}{\mu + \sum_{k \text{ s.t. } t_k < t_j} g_{kj}}.$$  (S18)

This choice of normalization factor relies on the implicit assumption that all potential triggering events are captured in the catalog. For this reason, it is essential to ensure completeness of the primary catalog. Events in the auxiliary catalog serve as potential triggering events, in particular of earthquakes at the beginning of the primary time period. They may not take the role of triggered events, and hence independence probabilities can only be defined for primary events. Completeness of the auxiliary catalog is beneficial, but not strictly required.

The sum of independence probabilities of events in the primary catalog yields the expected number of independent events, $n_{ind}$. Similarly, summing up the triggering probabilities of each triggering event, one obtains its expected number of aftershocks.

In the maximization step, the parameters are optimized to maximize the log likelihood of the observed data, assuming the expected number of independent events and the expected number of aftershocks of each event resulting from the preceding expectation step.

We stop the algorithm as soon as the cumulative absolute difference between parameters of two consecutive maximization steps falls below a threshold of 10-3. Table 1 in the paper shows the set of parameters obtained.

### ETAS Simulation

For the simulation of synthetic catalogs, we start by generating independent events. The number of independent events is drawn from a Poisson distribution. Its mean is calculated as the expected number of independent events for the region and time period in question. (Here, the parameter $\mu$ is used.) Time and location of each independent event are then drawn from uniform distributions. Especially for event location this is a massive generalization of reality. For the purpose of this analysis however, such a generalization will likely have a minor effect on the results. Clusters are detected depending on distances in-between events. If event locations are more evenly distributed, clusters might more easily be detected. The effect of declustering could therefore be more pronounced in synthetic catalogs compared to real catalogs. Since the purpose of using synthetic catalogs is to reproduce similar effects to those that are observed in real data, an amplification of the effects does not invalidate the argument.

Denoting independent events as events of generation 0, we recursively simulate events of subsequent generations until no more aftershocks are produced, or all aftershocks lie outside the relevant time window. For each event of generation $i$ that lies in the relevant time window, we

-   calculate its expected number of aftershocks as

$$n_{AS}(m, \Delta t_0, \Delta t_1) = k_0 \cdot e^{(a \cdot (m - m_c))} \cdot \frac{\pi}{\rho} \cdot \left( d \cdot e^{\gamma \cdot (m - m_c)} \right)^{-\rho} \cdot e^{c/\tau} \cdot \tau^{-\omega}$$

$$\cdot \left( \Gamma \left( -\omega, \frac{\Delta t_0 + c}{\tau} \right) - \Gamma \left( -\omega, \frac{\Delta t_1 + c}{\tau} \right) \right), \tag{S19}$$

where $\Gamma(s, x)$ is the upper incomplete gamma function, and $\Delta t_0, \Delta t_1$ are the positive time difference between the event and the start and end of the primary time period, respectively.

- draw its actual number of aftershocks from a Poisson distribution with mean rate $n_{AS}(m, \Delta t_0, \Delta t_1)$.

- simulate aftershock occurrence times by simulating inter-event time differences following the Omori law. For this, we simulate uniformly distributed random numbers between 0 and 1 and apply the inverse of the Omori law CDF with parameters $c, \omega, \tau$

- simulate aftershock locations isotropically around its epicenter by simulating inter-event distance and angle. Again, we simulate uniformly distributed random numbers between 0 and 1 and apply the inverse of the CDF induced by the spatial kernel with parameters $d, \gamma, \rho$, yielding the distance. The inter-event angle is sampled from a uniform distribution on the interval $[0, 2\pi)$.

- simulate aftershock magnitudes by simulating uniformly distributed random numbers and applying the inverse of the CDF induced by the GR-law with fixed $b$-value.

Because the time window is limited, more and more aftershocks will be simulated to lie outside the relevant period and hence the algorithm stops after having generated a finite number of earthquakes. Note that the time period is not required to be limited for the ETAS simulation to stop after a finite number of generations. One can calculate the so-called branching ratio

$$n = \int_{m_c}^{\infty} p(m) \cdot n_{AS}(m, 0, \infty) dm, \tag{S20}$$

where $p(m) = \beta \cdot e^{-\beta \cdot (m - m_c)}$ is the probability that an earthquake of magnitude above $m_c$ has magnitude $m$, and $\Gamma(-\omega, \infty) := 0$ in the calculation of $n_{AS}(m, 0, \infty)$. This branching ratio represents the average number of expected direct aftershocks of any earthquake. If $n < 1$, the sum of the geometric series of higher order aftershock numbers converges, and hence is finite. In other words, the ETAS process is in a subcritical regime. Therefore, the expected number of generations to be simulated before all events stop producing aftershocks, is also finite (Riziou et al., 2017). Based on the ETAS parameters obtained in the inversion, we find that $n \approx 0.89 < 1$, implying that simulations of aftershock chains will stop after finitely many generations.

Because the time delay between a triggering event and its aftershock can be in the order of months or even years, aftershock chains starting with independent events prior to the relevant time window are missing in such simulations. The characteristic waiting time $\tau$ is around 26.7 years in our case. Since aftershock chains can go on for many generations, we choose the start of our simulation period generously early on January 1, 1850, 120 years before the start of the auxiliary time period used for further analysis.

## ETAS Declustering

It has been proposed by Zhuang et al. (2002) that ETAS can also be used for declustering. We consider two versions of ETAS-based declustering, which differ in their definition of mainshocks. As was mentioned above, one component of ETAS inversion is to calculate probabilities $p_{ij}$ of each event $e_j$ to be triggered by each other event $e_i$, and the probability $p_i^{ind}$ of each event to be an independent event.

Regarding independent events to be mainshocks, one straightforward way to decluster is to weight each event in the catalog by its independence probability. The $b$-value can then be calculated using weighted magnitudes. We call this first version of ETAS declustering 'ETAS-Background'.

However, declustering methods normally distinguish between mainshocks and fore- or aftershocks, rather than independent events and triggered events in the ETAS sense. Going from independent and triggered events to mainshocks, fore- and aftershocks requires the extra step of identifying earthquake clusters and then imposing the rule of maximum magnitude to identify mainshocks. The probabilities $p_{ij}$ and $p_i^{ind}$ can be utilized to identify clusters, yielding an additional ETAS-based method of declustering.

We split our primary catalog into independent and triggered events as follows. Those events with the highest independence probabilities are defined to form the set $E_{ind}$ of independent events. The number of independent events is chosen to be $[n_{ind}]$, the closest integer to $n_{ind}$.

$$E_{ind} = \{e_i \mid p_i^{ind} \geq p_{thresh}\}, \tag{S21}$$

where $p_{thresh}$ is the maximum threshold such that $E_{ind}$ contains $[n_{ind}]$ events. Each independent event forms its own cluster. All remaining events are then chronologically added to one of the existing clusters. More precisely, each triggered event is added to the cluster with the highest responsibility for having triggered it. The responsibility $r_{C_i}$ for cluster $C_i$ for having caused event $e_j$ is given as the cumulative triggering probability of $C_i$,

$$r_{C_i} = \sum_{k \text{ s.t. } e_k \in C_i} p_{kj}. \tag{S22}$$

Events of the auxiliary catalog are allowed to have responsibility for a primary event. As there is no information on triggering probabilities $p_{ij}$ between auxiliary events, they are not seen as part of any cluster, which is why each auxiliary event is interpreted as a separate cluster. In this way, primary events may be allocated to a cluster originating in the auxiliary time period. These events are excluded from further analysis, since they are assumed to be incomplete.

The ETAS method for declustering does not depend on any input parameters. Unlike with other methods, the optimal parameters for declustering can be naturally obtained by calibrating the model on the data. 'ETAS-Main', in agreement with the other three methods, defines the largest event of each cluster to be the mainshock of the cluster.

Note that the non-parametric stochastic declustering algorithm proposed by Marsan and Lengline (2008) is not used here. This is because of its similarity to the already considered parametric stochastic declustering alternative provided by the ETAS model. The main difference to ETAS declustering is that the triggering rate, here denoted by $g(t, x, y, m)$, is there obtained empirically, without presuming the laws (S12)-(S14). In their analysis of southern California seismicity, they observe that their empirically derived triggering rate follows laws similar to those described in (S12)-(S14), which, likewise, were originally discovered empirically.

## Text S5.

### Sensitivity of $b$-Value to $m_c$

A sensitivity analysis of the $b$-value to the completeness magnitude $m_c$, represented in Figure S3, shows that the $b$-value decrease after declustering is an effect that is observed regardless of the choice of $m_c$. For the low value of $m_c = 3.2$, Reasenberg and Zaliapin declustered catalogs' $b$-values do not significantly differ from the full catalog $b$-value. However, completeness at this value is not certain, especially not after large events. As we increase the completeness threshold and therewith the certainty of dealing with a complete catalog, $b$-values quickly start to differ substantially from the full-catalog $b$-value. Note that the extent of the decrease is characteristic of each method.

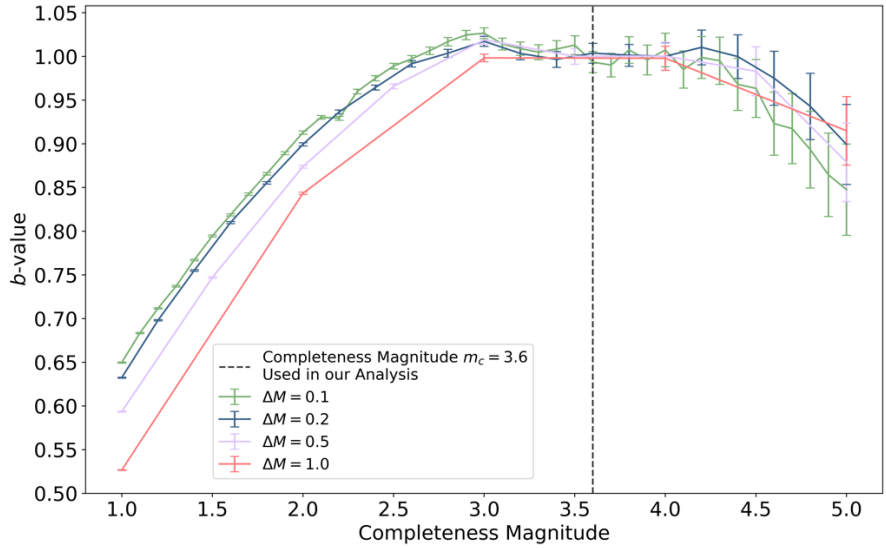

**Figure S1.** Sensitivity of $b$-value on completeness magnitude and magnitude bin size $\Delta M$.

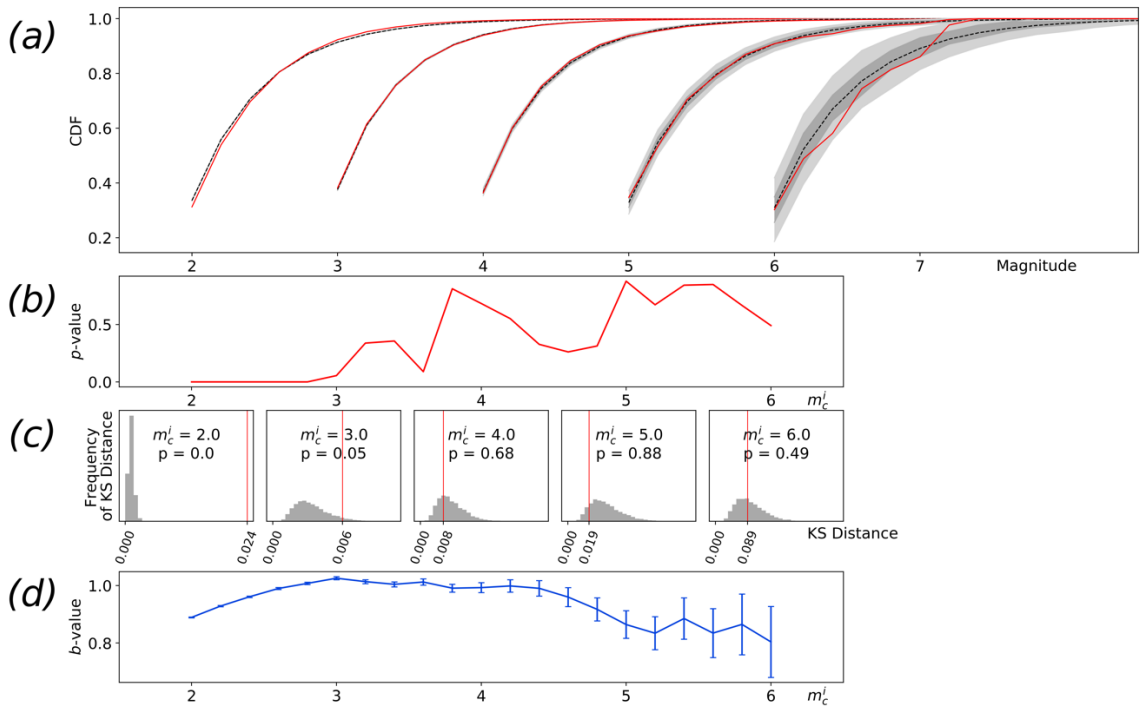

**Figure S2.** Testing $m_c^i$ between 2.0 and 6.0. (a): For integer values of $m_c^i$, theoretical and observed CDFs are represented as dotted black and red lines, respectively. Light grey and dark grey areas mark 5th to 95th, and 25th to 75th percentile of 10,000 randomly sampled CDFs. Grey areas are barely visible for small $m_c^i$, because they are in agreement with the theoretical CDF. (b): Evolution of $p$-value with $m_c^i$. (c): Derivation of $p$-value for integer values of $m_c^i$. Histogram shows the frequency distribution of KS distances of 10,000 randomly sampled CDFs to theoretical CDF. Red vertical line indicates KS distance of observed to theoretical CDF. (d) Evolution of $b$-value with $m_c^i$, with uncertainties.

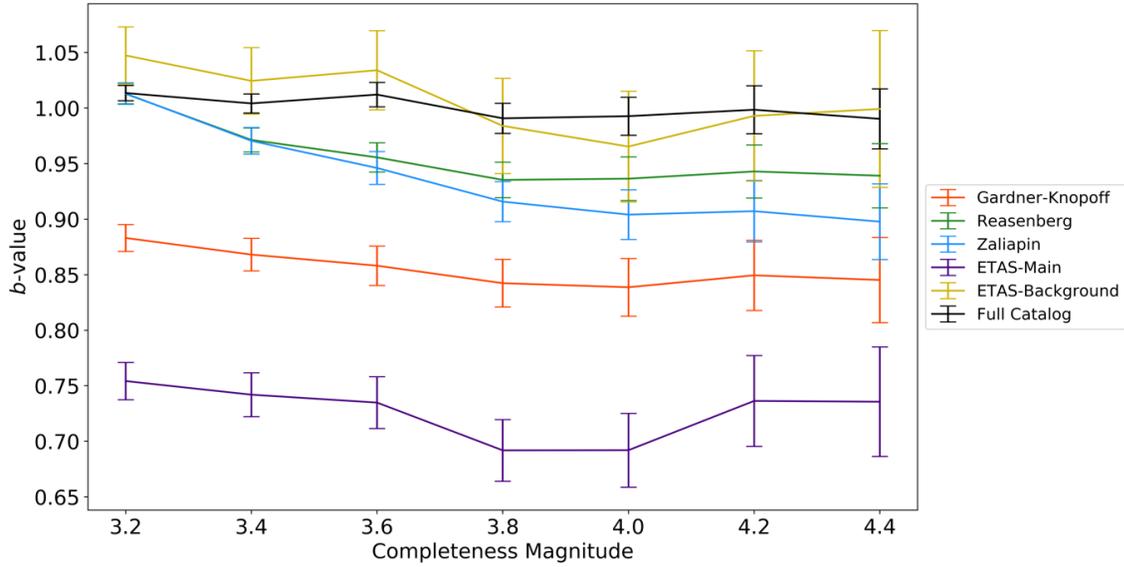

**Figure S3.** Sensitivity of mainshock $b$-value on completeness magnitude for different declustering methods.

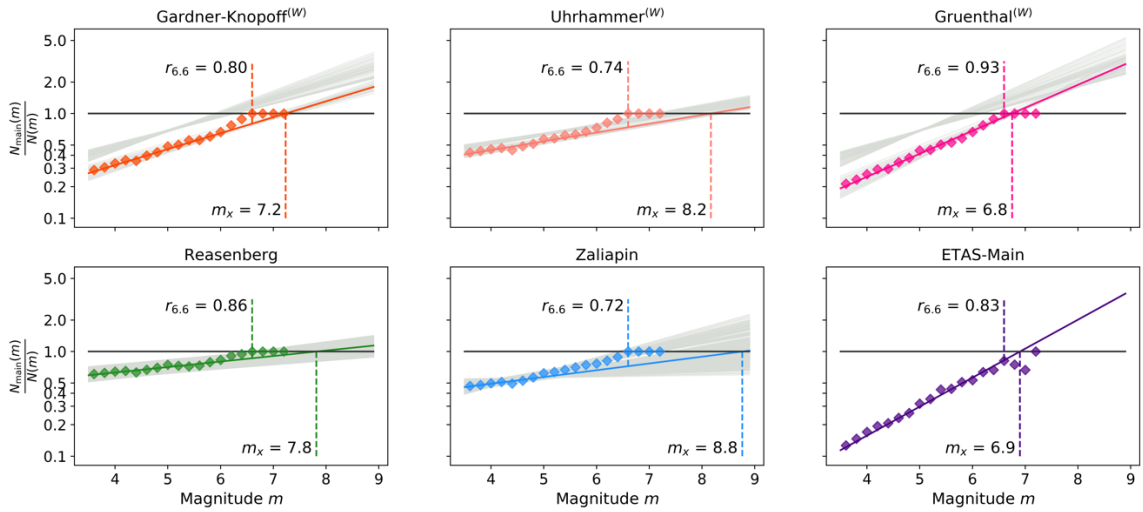

**Figure S4.** Observed and approximated (diamonds and lines) evolution of $r(m) = \frac{N_{main}(m)}{N(m)}$ for the California catalog. Declustering method varies with panel, marked with (W) are window methods. Grey lines show realizations of $r(m)$ for different declustering parameters, colored line shows the realization for standard parameters, black line marks $r(m) \equiv 1$. Colored diamonds mark observed values for standard declustering parameters. $m_x$ and $r_{6.6}$ are given for standard declustering parameters.

**Table S1.** Parameter ranges applied in the Reasenberg declustering algorithm. Standard parameter settings are given in bold characters.

| Parameter Name | Value Range |
|---|---|
| Interaction Formula | Reasenberg1985*, WellsCoppersmith1994* |
| $\tau_{min}$ | 0.5, **1.0**, 1.5, 2.0, 2.5 |
| $\tau_{max}$ | 3.0, 6.0, 8.0, **10.0**, 11.0, 13.0, 15.0 |
| $x_k$ | 0.0, 0.167, 0.333, **0.5**, 0.667, 0.833, 1.0 |
| $p$ | 0.75, 0.8 , 0.85, 0.9 , **0.95** |
| $r_{fact}$ | 5.0, **10.0**, 15.0, 20.0 |

**Table S2.** Parameter ranges and window types applied in the window declustering methods. Standard window and parameter settings are given in bold characters.

| Parameter Name | Value Range |
|---|---|
| Window Method | **GardnerKnopoff**, Gruenthal, Uhrhammer |
| $\tau_{max}$ | **None**, 15.0, 30.0 |
| $p_{\tau_{foreshock}}$ | 0.0, 0.1, 0.2, … , **1.0**, 1.1, 1.2, … , 1.9, 2.0 |

**Table S3.** Parameter ranges applied in the Zaliapin declustering algorithm. Standard parameter settings are given in bold characters.

| Parameter Name | Value Range |
|---|---|
| $d$ | 1.0, 1.1, 1.2, 1.3, **1.4**, 1.5, … , 2.4, 2.5 |
| $b$ | 0.0, 0.8, 0.85, 0.9, 0.95, **1.0**, 1.05, … , 1.45, 1.5 |
| $\theta$ | 0.8, 0.85, 0.9, 0.95, **1.0**, 1.05, 1.1, 1.15, 1.2 |